\documentclass[a4paper,11pt]{article}
\usepackage{pos}

\usepackage{subfigure}

\title{Dark sector search at BESIII}

\author*[a]{Zhi-Jun Li}
\author[a]{Zheng-Yun You}

\affiliation[a]{School of Physics, Sun Yat-sen University, Guangzhou 510275, China}

\emailAdd{lizhj37@mail2.sysu.edu.cn}
\emailAdd{youzhy5@mail.sysu.edu.cn}

\abstract{The Standard Model has achieved significant success in particle physics; however, there are still some unresolved puzzles beyond it. This motivates the exploration of the dark sector. We present the recent dark sector search results from BESIII experiment, which include the search for axion-like particles in the $J/\psi$ data set, the search for muon-philic particles in muon radiation, the search for massless dark photons in $D^0\to\omega\gamma'$ and $D^0\to\gamma\gamma'$, and the search for invisible decays of $K^0_S$.}

\FullConference{The 21st International Conference on Hadron Spectroscopy and Structure (HADRON2025)\\
 27 - 31 March, 2025\\
Osaka University, Japan\\}

\begin{document}
\maketitle

\section{Introduction}
The Standard Model (SM) has achieved significant success in identifying a series of "fundamental" particles, such as quarks, leptons, the photon, the Higgs boson, the $Z$ boson, the $W$ boson, and gluons, which have been extensively studied. However, there are still some unresolved puzzles, including dark matter (DM), the strong CP problem, the muon $g-2$ anomaly, the fermion mass hierarchy, matter-antimatter asymmetry, and more. These puzzles suggest the existence of the dark sector beyond the Standard Model, potentially containing 
new invisible particles and new particles decaying to SM particles.
The search for the dark sector provides an intriguing avenue for probing new physics (NP) beyond the SM. If the mass of dark sector particles falls within the MeV to GeV range, they can be accessed through high-intensity $e^+e^-$ collider experiments such as Beijing Spectrometer III (BESIII)~\cite{BESIII:2009fln}.

BESIII is a general-purpose spectrometer designed for the study of $\tau$-charm physics in the center-of-mass energy range from 2.0 to 4.7~GeV. It records symmetric $e^+e^-$ collisions provided by Beijing Electron Positron Collider II (BEPCII) storage ring~\cite{Yu:2016cof} and has collected large data samples in this energy region~\cite{BESIII:2020nme}, including 10 billion $J/\psi$ events, 2.7 billion $\psi(2S)$ events, $20~\rm{fb}^{-1}$ data sample at 3.773 GeV, and more than $20~\rm{fb}^{-1}$ data sample above 4.0 GeV in total. With the world's largest charmonium data samples, significant baryon and meson data samples can also be generated from charmonium decays. Leveraging these extensive data samples at BESIII and mature analytical techniques~\cite{Li:2024pox}, it becomes feasible to investigate the dark sector beyond the SM.

\section{Search for di-photon decays of an axion-like particle in radiative $J/\psi$ decays}

The QCD axion is predicted by the Peccei-Quinn solution to the strong CP problem, whose mass and coupling strength follow a linear relationship. The QCD axion is an excellent cold dark matter candidate; however, no signals have been observed to date. A simple extension of the QCD axion involves allowing arbitrary masses and coupling strengths, leading to what is called the axion-like particle (ALP). The ALP interacts with the SM photons through the operator
\begin{eqnarray}
\mathcal{L}=-\frac{1}{4}g_{a\gamma\gamma}aF^{\mu\nu}\tilde{F}_{\mu\nu},
\label{eq:ALP_operator}
\end{eqnarray}
where $g_{a\gamma\gamma}$ is the coupling strength between the ALP and the SM photons. The ALP can be produced from the decay of a heavy photon with the process $\gamma^* \to a\gamma$, and then decay via $a \to \gamma\gamma$ to be detected.

\begin{figure*}[htbp] \centering
	\setlength{\abovecaptionskip}{-1pt}
	\setlength{\belowcaptionskip}{10pt}

        \subfigure[]
        {\includegraphics[width=0.49\textwidth]{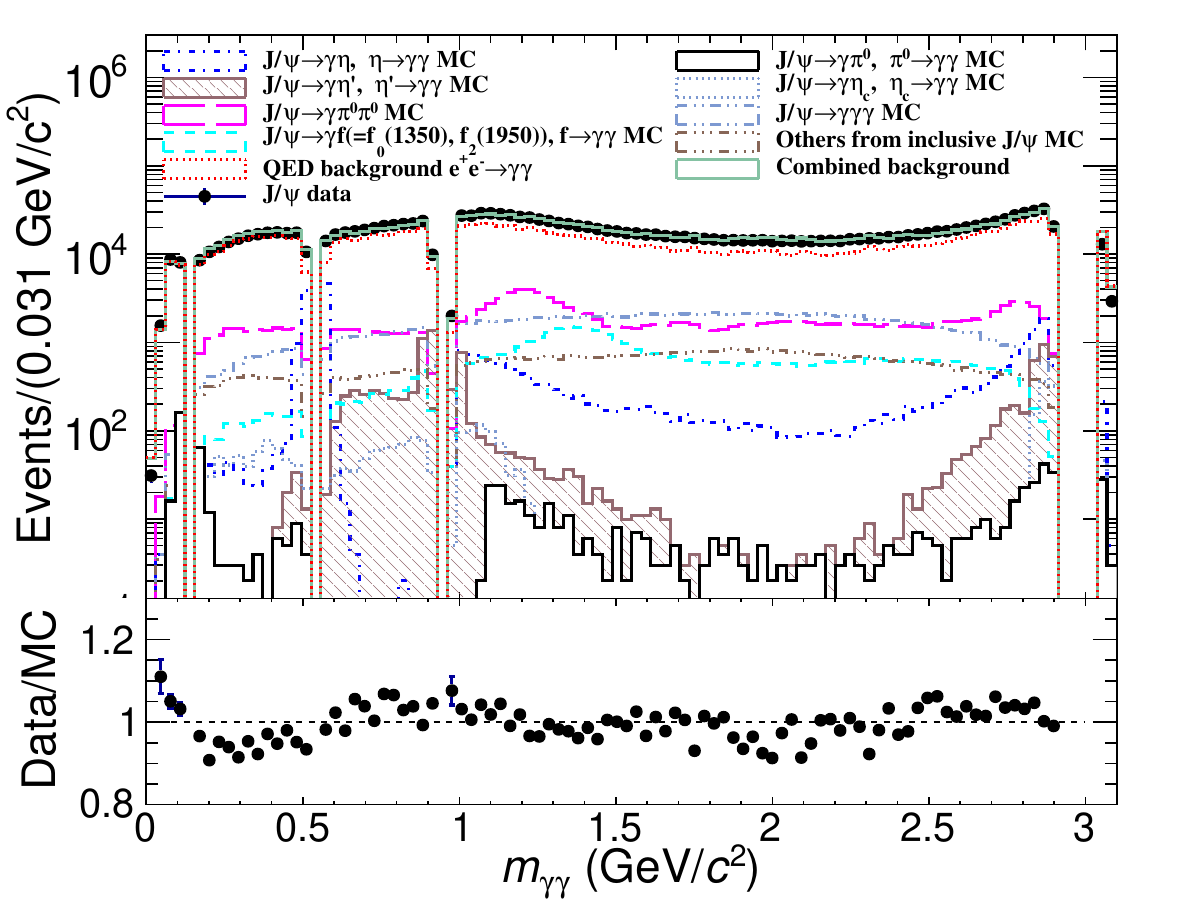}}
        \subfigure[]
        {\includegraphics[width=0.49\textwidth]{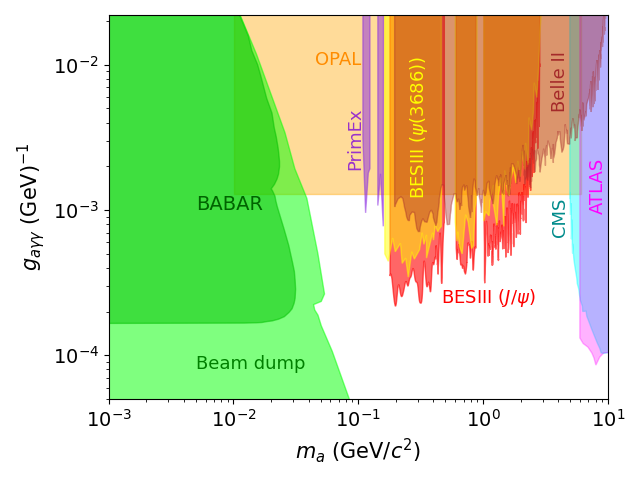}}
        
	\caption{
        (a) The distribution of the di-photon invariant mass in the ALP search. The black points represent the data samples, while the other histogram represents different background processes. (b) The constraints on the ALP-photon coupling. The red-filled regions indicate the excluded parameter space from the latest BESIII results, whereas the other filled regions correspond to previously excluded parameter spaces.
        } 
	\label{fig:ALP}
\end{figure*}

The main heavy photon source at BESIII is the $J/\psi$ events from $e^+e^-$ collisions, and the ALP can be produced from the decay of $J/\psi \to \gamma a$. The decay width of the ALP is given by $\Gamma_a = \frac{g^2_{a\gamma\gamma} m^3_a}{64\pi}$. Taking $g_{a\gamma\gamma} \sim 10^{-4}~\rm{GeV}^{-1}$ and $m_a \sim \rm{GeV}/c^2$, it results in the ALP decaying to two photons near the interaction point at BESIII. Utilizing $(10084 \pm 44) \times 10^6~J/\psi$ events collected at BESIII, we search for ALPs in the process of $J/\psi \to \gamma a \to \gamma \gamma \gamma$. Three photons are detected in the final states, and the ALP signal exhibits a peak in the $\gamma-\gamma$ invariant mass distribution of Figure~\ref{fig:ALP} (a). To reduce the background level, events around $\pi^0$, $\eta$, $\eta'$, and $\eta_c$ in the $\gamma-\gamma$ invariant mass distribution are excluded. By scanning the ALP mass hypothesis from $0.18 - 2.85~\rm{GeV}/c^2$, the maximum global signal significance is found to be 1.6 $\sigma$ at $M_a = 2786~\rm{MeV}/\it{c}^{\rm{2}}$, with no significant ALP signals discovered. The corresponding upper limits (ULs) on the branching fractions (BFs) for $J/\psi \to \gamma a \to \gamma \gamma \gamma$ are determined to be between $(3.6 - 53.1) \times 10^{-8}$ at the 95\% confidence level (C.L.)~\cite{BESIII:2024hdv}.
The BF of $J/\psi$ is sensitive to the coupling strength of $g_{a\gamma\gamma}$, with the relationship of~\cite{Merlo:2019anv}
\begin{eqnarray}
\frac{\mathcal{B}(J/\psi\to\gamma a)}{\mathcal{B}(J/\psi\to e^+e^-)}=\frac{m^2_{J/\psi}}{32\pi\alpha} g^2_{a\gamma\gamma} \left(  1-\frac{m^2_a}{m^2_{J/\psi}}  \right)^3.
\label{eq:ALP_BF}
\end{eqnarray}
Here, the coupling between the ALP ($a$) and $c$ quarks is disregarded. Assuming a BF of $\mathcal{B}(a \to \gamma\gamma)=100\%$, the ULs on the ALP-photon coupling strength are set to be $(2.2 - 101.8) \times 10^{-4}~\rm{GeV}^{-1}$ for $0.18 < m_a < 2.85~\rm{GeV}/c^2$, as shown in Figure~\ref{fig:ALP} (b), which represent the most stringent limits to date in this mass region. 
Note that if the BF $\mathcal{B}(a \to \gamma\gamma)$ is not 100\%, the constraints from all the experiments (not only BESIII) shown in Figure~\ref{fig:ALP} (b) will be less stringent.

\vspace{-0.0cm}
\vspace{-0.0cm}

\section{Search for a muon-philic scalar $X_0$ or vector $X_1$ via $J/\psi\to\mu^+\mu^-+\rm{invisible}$ decays}
The puzzle of the $g_{\mu}-2$ anomaly has garnered considerable attention in particle physics. One proposed solution is to introduce a new particle that couples specifically to the muon, thereby contributing to an additional effective coupling between the muon and the SM photons. In previous studies, most experiments have focused on new particles (such as dark photons) that couple with light fermions, such as light quarks or electrons, extending the constraints to the muon based on the assumption of coupling universality. However, if the new particle is exclusively muon-philic, the previous constraints derived from experiments involving light quarks may not be applicable.
A $U(1)_{\mathcal{L}_{\mu}-\mathcal{L}_{\tau}}$ model predicts the existence of a new massive scalar boson $X_0$ or a vector boson $X_1$ that only couples to the second and third generations of leptons ($\mu,~\tau,~\nu_{\mu},~\nu_{\tau}$) via the operators $\mathcal{L}_{\mu}^{\rm{scalar}} = -g'_0 X_0 \bar{\mu} \mu$ and $\mathcal{L}_{\mu}^{\rm{vector}} = -g'_1 X_1 \bar{\mu} \gamma^{\alpha} \mu$, where $g'_{0,1}$ denotes the coupling strengths. These light muon-philic scalar or vector particles can contribute to the muon anomalous magnetic moment and thereby help explain the $(g-2)_{\mu}$ anomaly~\cite{Cvetic:2020vkk}.

The main source of muons at BESIII arises from the decay $J/\psi \to \mu^+\mu^-$, and the muon-philic particles can be produced via the process $J/\psi \to \mu^+\mu^- X_{0,1}$, where $X_{0,1}$ is invisible in the search. Based on $(8.998 \pm 0.039) \times 10^9$ $J/\psi$ events, we perform a search for the muon-philic particles $X_{0,1}$ through the process $J/\psi \to \mu^+\mu^- X_{0,1}$. Three cases of muon-philic particles are considered.
The first case is referred to as the "vanilla" model, which assumes $m_{\chi} > m_{X_1}/2$ ($\chi$ is another light dark sector particle, such as a dark fermion, which could be the DM candidate), where the muon-philic vector particle $X_1$ can only decay to SM particle pairs. The decay channel $X_1 \to \nu\nu$ can become accessible in the invisible final state, with a BF $\mathcal{B}(X_1 \to \nu\nu) = 33\% \sim 100\%$ depending on the mass of $X_1$.
The second case is termed the "invisible" model, in which $m_{\chi} < m_{X_1}/2$, and the muon-philic vector particle $X_1$ primarily decays into an invisible DM particle pair, assuming the coupling strength $g'_D \gg g'_1$.
The third case is known as the "scalar" model, which assumes that the muon-philic scalar particle $X_0$ is long-lived or only decays to invisible final states.

\begin{figure*}[htbp] \centering
	\setlength{\abovecaptionskip}{-1pt}
	\setlength{\belowcaptionskip}{10pt}

        \subfigure[]
        {\includegraphics[width=0.32\textwidth]{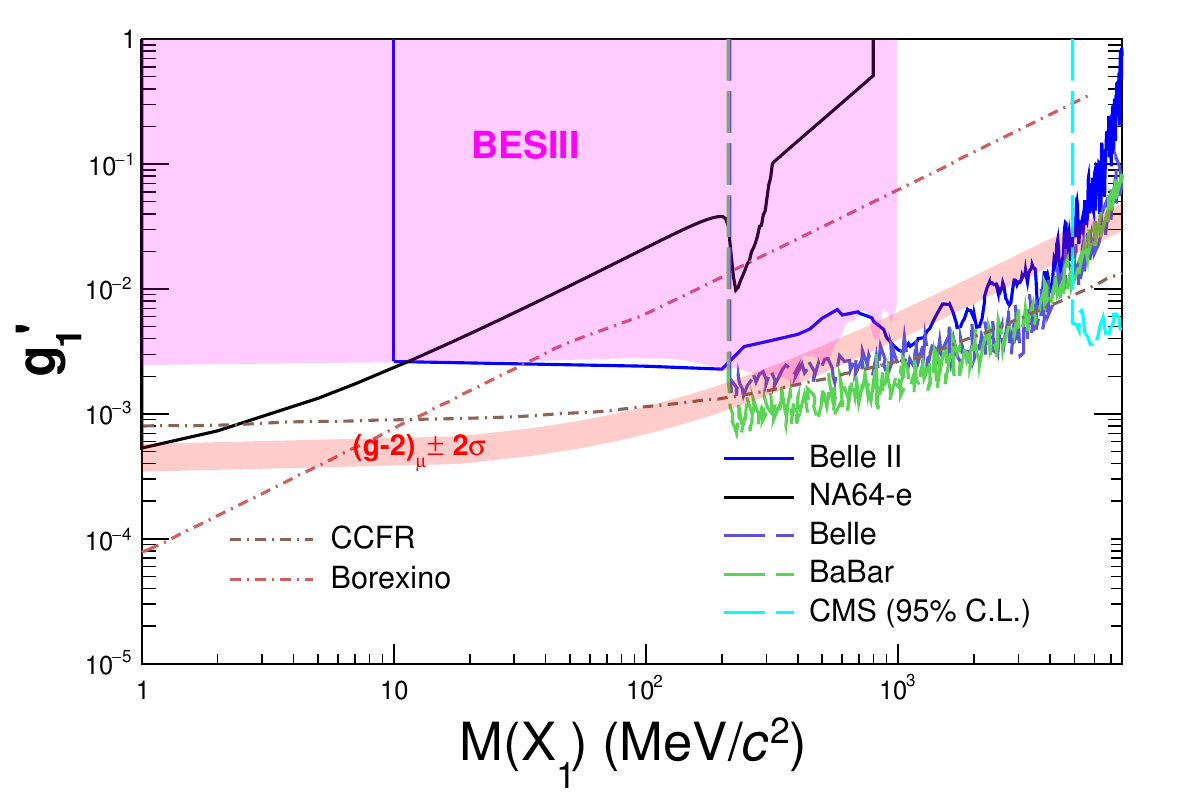}}
        \subfigure[]
        {\includegraphics[width=0.32\textwidth]{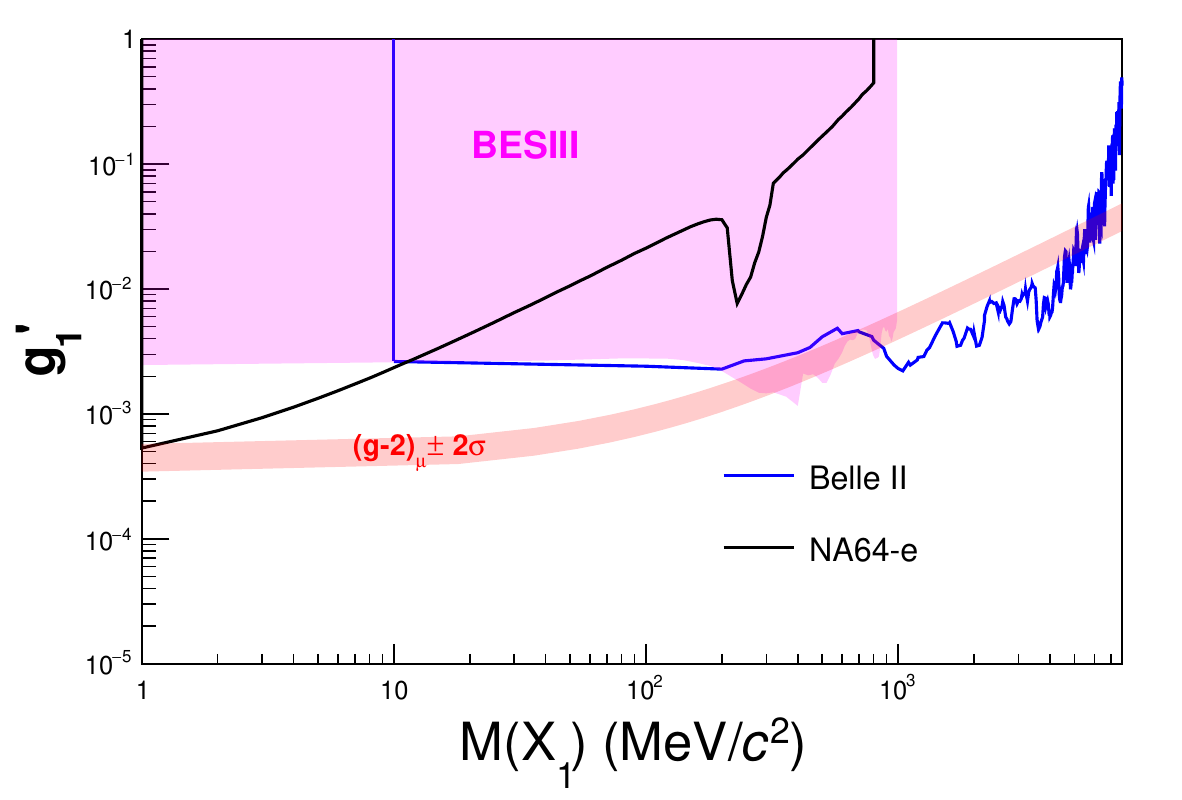}}
        \subfigure[]
        {\includegraphics[width=0.32\textwidth]{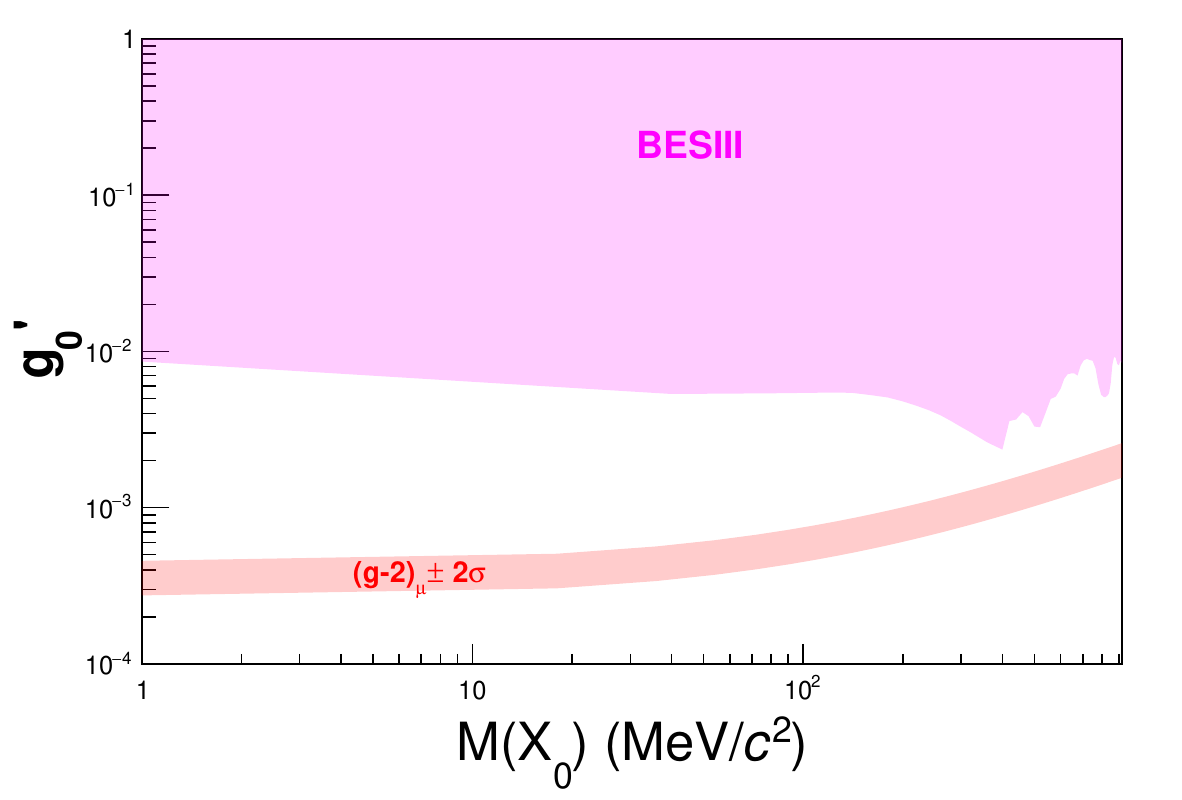}}\\
        
	\caption{
        The constraint of the coupling between muonphilic particle and the SM muons, where (a) is the ``vanilla" model, (b) is the ``invisible" model, and (c) is the ``scalar" model. The purple-red filled region represents the excluded parameter space from BESIII, while the light-red filled region indicates the parameter space that could explain the $g_{\mu}-2$ anomaly before 2025. The other lines depict the excluded parameter space from various other experiments.
        } 
	\label{fig:muonphilic}
\end{figure*}

The invisible muon-philic particle is reconstructed from the recoiling of $\mu^+\mu^-$, and the maximum local signal significance observed is 2.5 $\sigma$ at $m(X_{0,1}) = 720~\rm{MeV}/c^2$. No evidence for signals from the invisible decays of $X_{0,1}$ has been observed, and the coupling constraint at 90\% C.L. is presented~\cite{BESIII:2023jji} in Figure~\ref{fig:muonphilic}. 
For the "vanilla" model, results from BaBar, CMS, and Belle are based on the process $X_1 \to \mu^+\mu^-$, while Belle II and BESIII examine the decay $X_1 \to \nu\bar{\nu}$. For the "invisible" model, BESIII provides better sensitivity in the mass range of 200 to 860 $\rm{MeV}/c^2$. In the case of the "scalar" model, BESIII presents the first constraint for the invisible scalar particle $X_0$.
As shown in Figure~\ref{fig:muonphilic}, most of the parameter space to explain the $g_{\mu}-2$ anomaly cannot be achieved with the current experimental set. However, it is worth noting that when the author wrote the proceeding, a new lattice QCD calculation had been reported that is consistent with the experimental value of $g_{\mu}-2$~\cite{Aliberti:2025beg}. If the lattice QCD calculation is accurate, the allowed parameter space from $g_{\mu}-2$ will be changed to a lower UL.
\vspace{-0.0cm}
\vspace{-0.0cm}

\section{Search for massless dark photon in $D^0\to\omega\gamma'$ and $D^0\to\gamma\gamma'$ decays}
The dark photon ($\gamma'$) is introduced in the minimal extension of the SM through an additional Abelian gauge group. If the symmetry of this additional Abelian gauge group is spontaneously broken, the dark photon will acquire mass and coupling to SM fermions via kinetic mixing. Conversely, if the symmetry remains unbroken, the dark photon is massless and does not have any direct coupling with SM particles from a theoretical perspective.
The massless dark photon plays a significant role in the dark sector, such as providing a new long-range force for DM, serving as a potential solution to the fermion mass hierarchy, explaining the excess in the decay process $B^+ \to K^+ \nu\bar{\nu}$, elucidating the origin of the CKM matrix structure, and addressing the vacuum instability problem in the SM Higgs sector~\cite{Fabbrichesi:2020wbt}.

Compared to the stringent constraints on the massive dark photon, the massless dark photon remains significantly less constrained. Searches for the massless dark photon can only be conducted through higher-dimensional operators, such as~\cite{Dobrescu:2004wz}
\begin{eqnarray}
\mathcal{L}_{\rm{NP}} = & \frac{1}{\Lambda^2_{\rm{NP}}} \left( C^u_{jk} \bar{q}_j \sigma^{\mu\nu} u_k \tilde{H} + C^d_{jk} \bar{q}_j \sigma^{\mu\nu} d_k H + C^l_{jk} \bar{l}_j \sigma^{\mu\nu} e_k H + h.c. \right) F'_{\mu\nu},
\label{eq:dimension-six operator}
\end{eqnarray}
where $\Lambda_{\rm{NP}}$ is the NP energy scale and $C_{jk}$ are the dimensionless coefficients. This operator inherently includes flavor-violating couplings, such as the $cu \gamma'$ coupling.
Based on $7.9~\rm{fb}^{-1}$ of $e^+e^-$ annihilation data at $\sqrt{s} = 3.773$ GeV, we search for the massless dark photon in the decay processes of $D^0 \to \omega \gamma'$ and $D^0 \to \gamma \gamma'$. Since the $D^0$ mesons are always produced in pairs at $\sqrt{s} = 3.773$ GeV, one side of $D^0$ can be tagged using some well-known SM hadronic channels. We then search for the NP decays on the other side, a technique known as the double tag method.

\begin{figure*}[htbp] \centering
	\setlength{\abovecaptionskip}{-1pt}
	\setlength{\belowcaptionskip}{10pt}

        \subfigure[]
        {\includegraphics[width=0.32\textwidth]{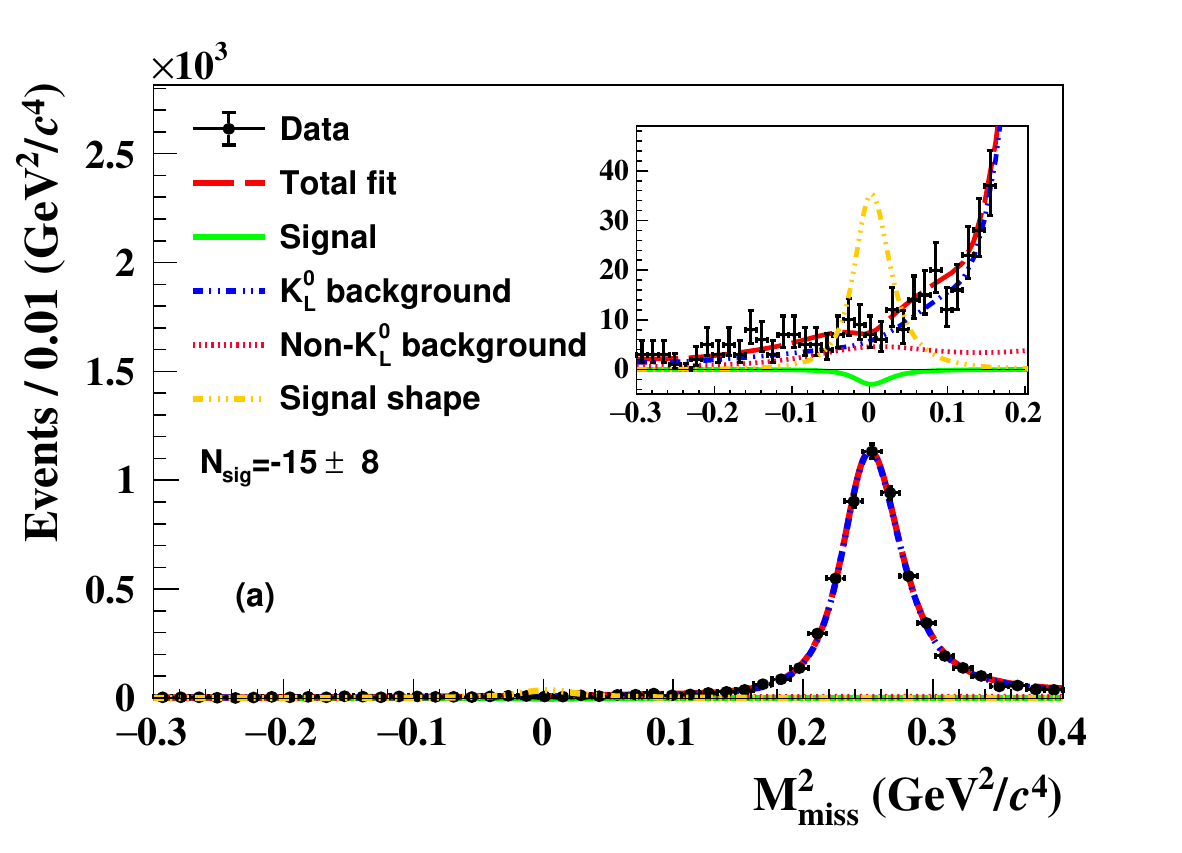}}
        \subfigure[]
        {\includegraphics[width=0.32\textwidth]{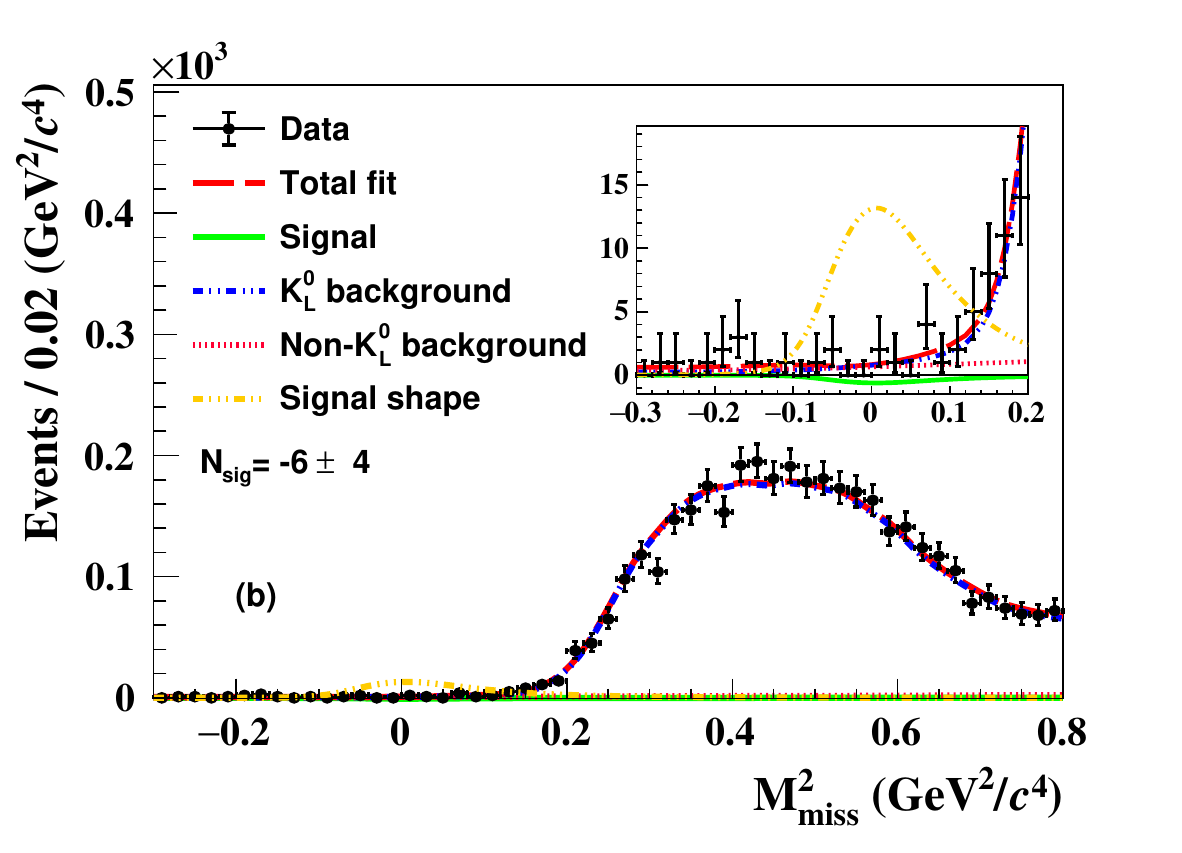}}
        \subfigure[]
        {\includegraphics[width=0.32\textwidth]{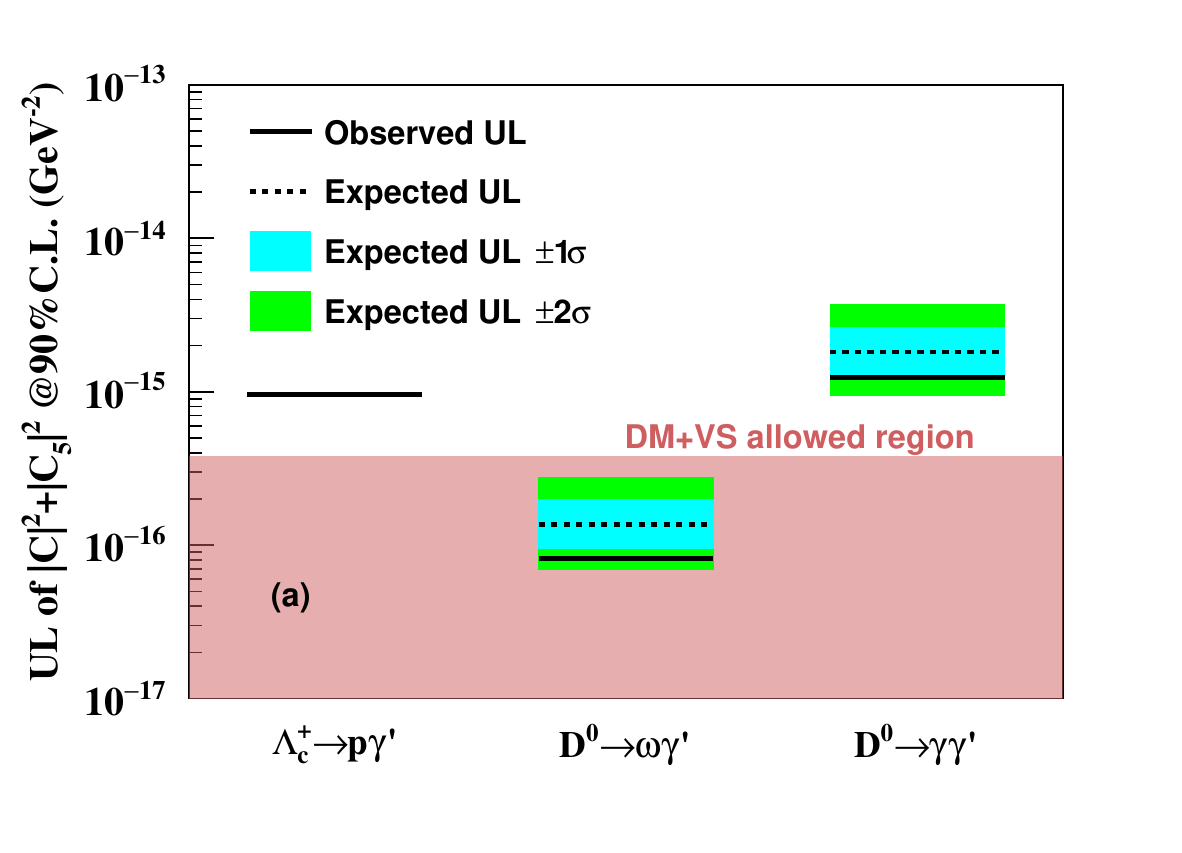}}\\
        
	\caption{
        (a) (b) The $M^2_{\rm{miss}}$ distribution of the candidates for $D^0 \to \omega \gamma'$ (a) and $D^0 \to \gamma \gamma'$ (b). The black points represent the data samples, while the other lines correspond to different components. (c) The constraint on the $cu \gamma'$ coupling. The black solid lines indicate the observed ULs from different processes, and the red filled regions represent the allowed regions from DM and VS.
        } 
	\label{fig:Gp}
\end{figure*}

The invisible massless dark photon can be reconstructed from the recoil of visible SM particles, as illustrated in Figure~\ref{fig:Gp} (a) and (b). No significant signal has been observed in the data samples, and the ULs at the 90\% C.L. are determined to be $1.1 \times 10^{-5}$ and $2.0 \times 10^{-6}$ for the decay processes $D^0 \to \omega \gamma'$ and $D^0 \to \gamma \gamma'$, respectively~\cite{BESIII:2024rkp}.
\vspace{-0.0cm}
\vspace{-0.0cm}
The BF of $D^0$ decaying to a massless dark photon is highly sensitive to the NP energy scale and dimensionless coefficients in Eq.~\ref{eq:dimension-six operator}, with the relationship given by~\cite{Su:2020yze}
\begin{eqnarray}
\mathcal{B}(D^0 \to V \gamma') = \frac{\tau_D f^2_{DV} (m^2_D - m^2_V)^3}{2\pi m^3_D} (|\mathcal{C}|^2 + |\mathcal{C}_5|^2),
\end{eqnarray}
and 
\begin{eqnarray}
\mathcal{B}(D^0 \to \gamma \gamma') = \frac{\alpha_e}{2} \tau_D f^2_{D\gamma} m^3_{D} (|\mathcal{C}|^2 + |\mathcal{C}_5|^2),
\end{eqnarray}
where $f$ is the form factor of the $D$ decay, $\mathcal{C} = \Lambda_{\rm{NP}}^{-2} (C^u_{12} + C^{u*}_{12}) \nu / \sqrt{8}$, $\mathcal{C}_5 = \Lambda_{\rm{NP}}^{-2} (C^u_{12} - C^{u*}_{12}) \nu / \sqrt{8}$, and $\nu$ is the expected value of the Higgs vacuum.
For the decay $D^0 \to \omega \gamma'$, it provides a stringent constraint on the NP energy scale-related parameters with $|\mathcal{C}|^2 + |\mathcal{C}_5|^2 < 8.2 \times 10^{-17} \rm{GeV}^{-2}$. This new constraint represents an improvement of more than one order of magnitude and enters the allowed regions for DM and vacuum stability (VS) for the first time, as shown in Figure~\ref{fig:Gp} (c). Conversely, the constraint from $D^0 \to \gamma \gamma'$, despite having a better UL on the BF, is weaker due to the additional factor of $\alpha_e$ in the decay.

\section{Search for $K^0_S$ invisible decays}
The BF of the $K^0_S$ invisible decay is extremely low due to loop diagrams and helicity suppression. However, some NP beyond the SM may enhance its BF, such as considering $K^0_S$ decaying to a DM pair or considering ordinary-mirror particle oscillations, which could increase the invisible BF to $10^{-6}$. 
Furthermore, the invisible decay of $K^0_S$ can also provide valuable input for tests of the CPT (Charge, Parity, and Time reversal) symmetry. This is because the Bell-Steinberger relation connects CPT violation to the amplitudes of all decay channels of neutral kaons, albeit under the assumption that no invisible modes exist currently~\cite{Gninenko:2014sxa}.

\vspace{-0.0cm}
\begin{figure*}[htbp] \centering
	\setlength{\abovecaptionskip}{-1pt}
	\setlength{\belowcaptionskip}{10pt}

        \subfigure[]
        {\includegraphics[width=0.6\textwidth]{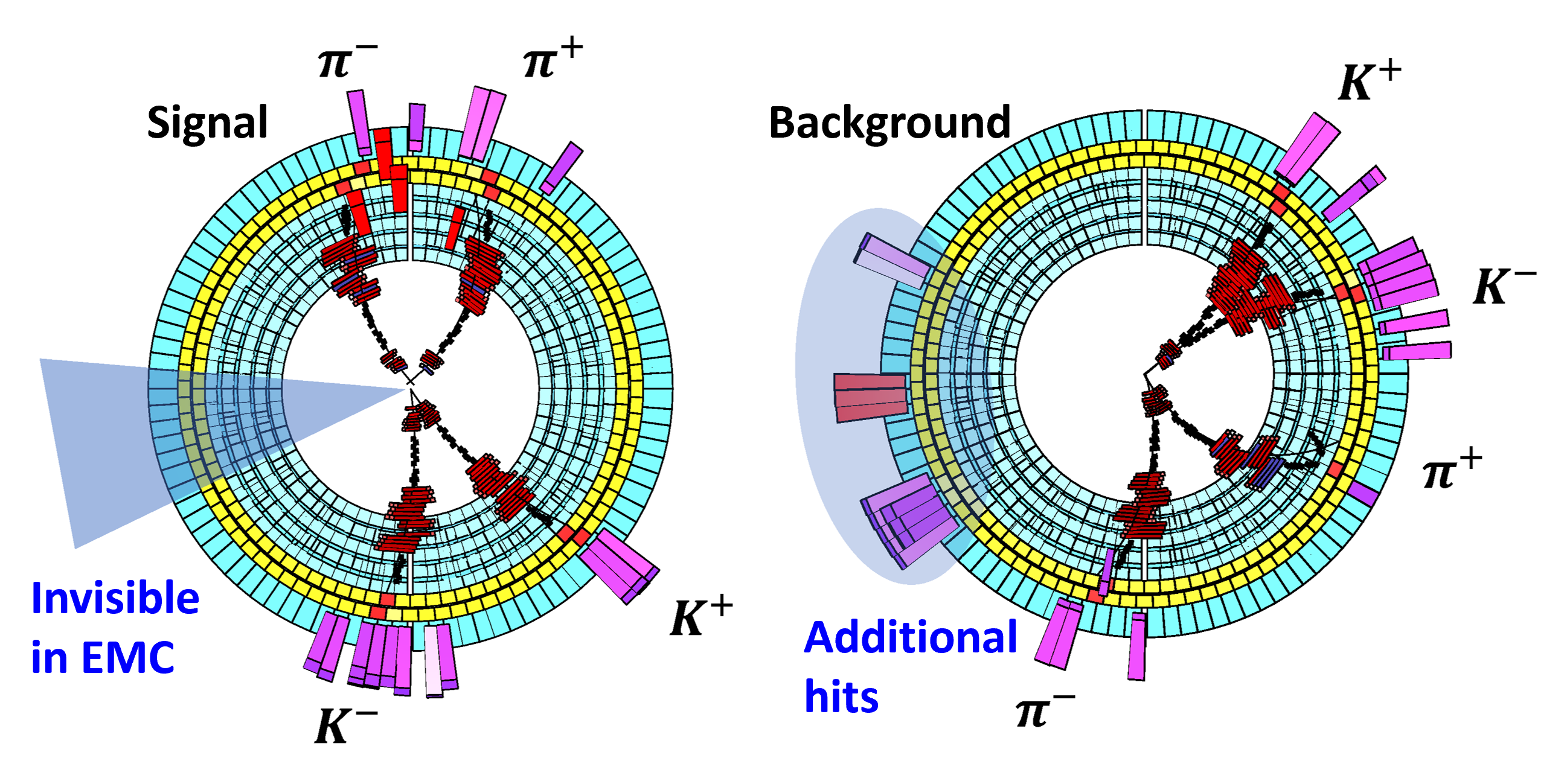}}
        \subfigure[]
        {\includegraphics[width=0.38\textwidth]{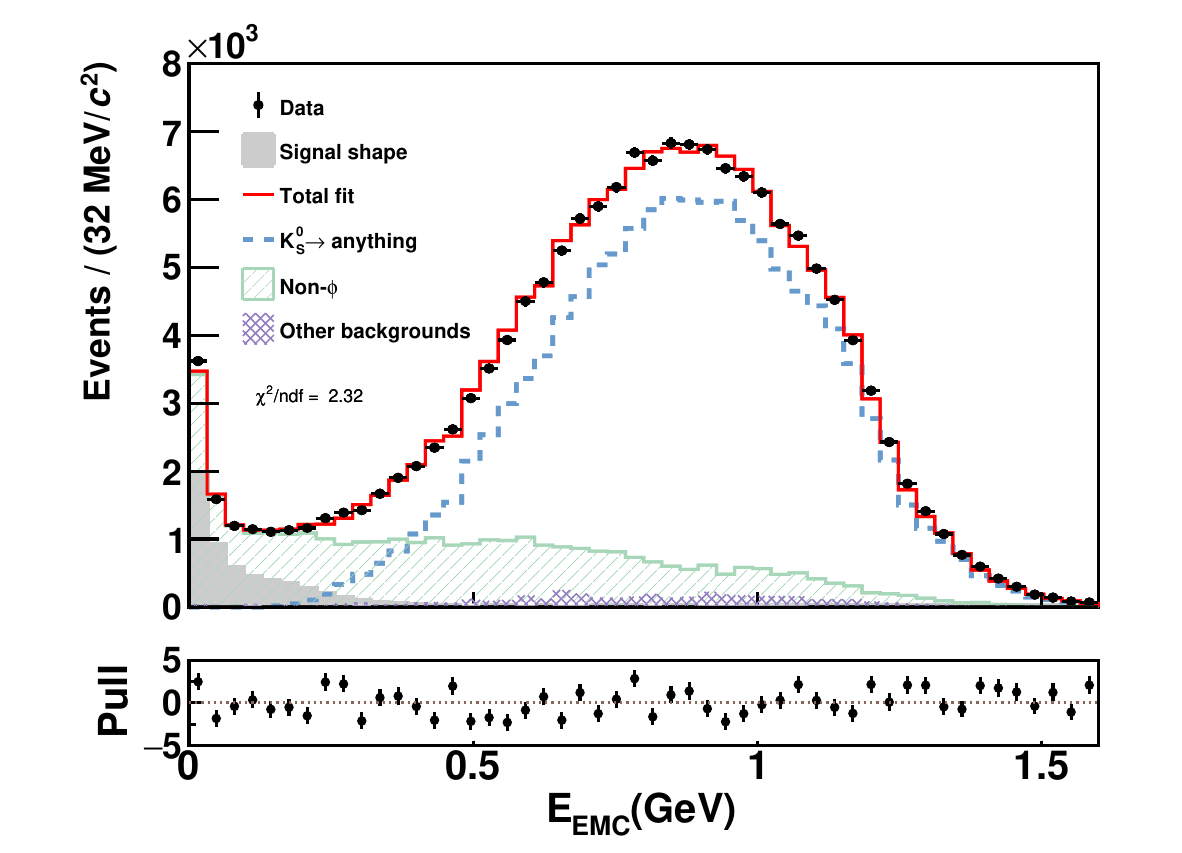}}\\
        
	\caption{
        (a) The event display for the $K^0_S\to\rm{invisible}$ signal process (left) and the background process (right) from the Monte Carlo simulation. The light blue barrel represents the EMC sub-detector, while the purple blocks indicate the hits in the EMC. 
        (b) The distribution of the deposited energy for the accepted candidates of $K^0_S\to\rm{invisible}$ analysis. The black points represent the data samples, and the other lines or histograms correspond to different components.
        } 
	\label{fig:Ks}
\end{figure*}
\vspace{-0.0cm}

The first direct search for $K^0_S$ invisible decay is conducted at BESIII using 10 billion $J/\psi$ events. The $K^0_S$ source is obtained from the decay $J/\psi \to \phi K^0_S K^0_S$, which features a relatively lower background level due to the prohibition of the decay $J/\psi \to \phi K^0_S K^0_L$ by C parity conservation. In our search, the $\phi$ meson is reconstructed via the decay $\phi \to K^+K^-$, one of the $K^0_S$ mesons is reconstructed through the decay $K^0_S \to \pi^+\pi^-$, while another $K^0_S$ meson is utilized to search for the invisible decay.
The invisible signal is identified using the sub-detector of the Electron Magnetic Calorimeter (EMC). For the invisible signal, there are no additional hits recorded in the EMC; conversely, for the SM background, there will be some additional hits, as demonstrated in Figure~\ref{fig:Ks} (a). Therefore, the deposited energy in the EMC can be utilized to extract the signal yield, as illustrated in Figure~\ref{fig:Ks} (b). 
In the deposited energy distribution, the peak at zero corresponds to the $J/\psi \to K^+ K^- K^0_S K^0_L$ background when there is no $\phi$ present, which is characterized by the $\phi$ sideband region. No significant signals are found exceeding the background, and the UL on the BF of the $K^0_S$ invisible decay is set to be $8.4 \times 10^{-4}$ at 90\% C.L.~\cite{BESIII:2025kjj}. This represents the first direct measurement of $K^0_S \to \text{invisible}$; however, the UL remains above the predictions from NP.

\section*{Acknowledgements}
This work is supported by the National Key R\&D Program of China under Contracts Nos. 2023YFA1606000; National Natural Science Foundation of China~(Grant No. 12175321, 11975021).

\end{document}